\documentclass{PoS}
\usepackage{lineno}

\title{Coherent J/$\psi$ photoproduction in Pb-Pb collisions with nuclear overlap studied with ALICE at the LHC}

\ShortTitle{Coherent J/$\psi$ photoproduction in peripheral Pb-Pb collisions}

\author{\speaker{L. Massacrier for the ALICE Collaboration} \\
        IPNO, CNRS-IN2P3, Univ. Paris-Sud, Universit\'e Paris-Saclay, 91406 Orsay Cedex, France\\
        E-mail: \email{massacrier@ipno.in2p3.fr}}

\abstract{

The photoproduction of heavy vector mesons in the electromagnetic interactions of ultra-relativistic nuclei is sensitive to the gluon distribution in the nucleus and thus to cold nuclear matter effects like shadowing or parton saturation. Besides the well known observations of vector meson production in ultra-peripheral collisions,  first observations of an excess over the expected hadronic J/$\psi$ production at very low transverse momentum ($p_T < $~0.3 GeV/$c$) in peripheral and semi-central nucleus-nucleus collisions both at LHC and RHIC energies were interpreted as the first sign of coherent J/$\psi$ photoproduction occurring in Pb-Pb collisions with nuclear overlap. The ALICE Collaboration published the J/$\psi$  coherent photoproduction cross sections in peripheral and semi-central Pb-Pb collisions at $\sqrt{s_{\rm NN}}$ = 2.76~TeV and forward rapidity ($2.5<y<4.0$). Using the LHC Run-2 data, ALICE presents preliminary results in peripheral Pb-Pb collisions at $\sqrt{s_{\rm NN}}$ = 5.02 TeV at mid-rapidity ($|y|<0.9$) and forward rapidity. Thanks to the very good tracking resolution of the central barrel, the extraction of the $p_T$-differential cross section was also possible, strengthening the photoproduction origin of the observed J/$\psi$ excess. The quantitative understanding of this low-$p_T$ excess poses significant theoretical challenges since the J/$\psi$ photoproduction depends on the collision dynamics as well as on the photon-flux and the photonuclear cross section. In this proceeding, we present the latest ALICE measurements on J/$\psi$ photoproduction cross section in peripheral Pb-Pb collisions, with emphasis on the new forward measurement in the dimuon decay channel at $\sqrt{s_{\rm NN}}$ = 5.02~TeV. These results will be discussed and compared to several model calculations of J/$\psi$ photoproduction in Pb-Pb collisions with nuclear overlap.}

\FullConference{International Conference on Hard and Electromagnetic Probes of High-Energy Nuclear Collisions\\
		30 September - 5 October 2018\\
		Aix-Les-Bains, Savoie, France}

\begin{document}


\small

The physics of Ultra-Peripheral Collisions (UPC) is extensively reviewed in \cite{Baltz:2007kq, Bertulani:2005ru}. In such collisions, in which the nuclei are separated by an impact parameter larger than the sum of their radii, two-photon and photonuclear interactions can be studied, since hadronic interactions are strongly suppressed and the cross section for photon-induced reactions is large due to the strong electromagnetic field of the nuclei. Exclusive photoproduction of vector mesons, characterised by a clean final-state with no other produced particles except the vector meson, is of particular interest in heavy-ion collisions. Indeed it allows one to probe the nuclear gluon distribution functions \cite{Rebyakova:2011vf}, which are poorly known at low Bjorken-$x$ \cite{Eskola:2009uj}. Recently, the ALICE Collaboration reported the first measurement of an excess at very low transverse momentum ($p_{\rm T}$) in the yield of J/$\psi$ charmonium with respect to expectations from hadroproduction, in peripheral Pb-Pb collisions \cite{Adam:2015gba}. This excess was interpreted as a demonstration of coherent J/$\psi$ photoproduction, occurring in Pb-Pb collisions with nuclear overlap. In such coherent process, quasi-real photons coherently produced by the strong electromagnetic field of the first lead nuclei, interact also coherently with the whole second nuclei, to produce a charmonium. Theoretically, coherent photoproduction of vector meson when the nuclei interact hadronically raises several questions; for instance on how the coherence condition can survive when the nuclei are broken during the hadronic interaction. The first ALICE measurement was followed by a similar observation of an excess in the J/$\psi$ yield at very low $p_{\rm T}$ by the STAR Collaboration in Au-Au and U-U collisions at lower collision energies \cite{Zha:2018ohg}. In this proceeding, we present the results from the 2015 Pb-Pb data taken by ALICE at $\sqrt{s_{\rm NN}}$ = 5.02 TeV, both at mid and forward rapidity, as well as comparisons to several theoretical models.  \\
\indent{The mid-rapidity analysis ($\mid y \mid <$ 0.9) is performed in the dielectron decay channel on a dataset corresponding to an integrated luminosity of $\cal L_{\rm int} \sim$ 10 $ \mu$b$^{-1}$, while the forward analysis (2.5~$< y <$~4) is done in the dimuon decay channel on a dataset corresponding to an integrated luminosity of $\cal L_{\rm int} \sim$~220~$ \mu$b$^{-1}$. The ALICE detector is described in \cite{Aamodt:2008zz}. The detectors used in both analyses are the V0 hodoscopes,  the Zero Degree Calorimeters (ZDC) and the Inner Tracking System (ITS). The V0 is used for triggering and the determination of the collision centrality. Together with the ZDC timing information, the V0 is also used to reduce the beam induced background. A minimum energy deposition is also requested in the part of the ZDC that detects neutrons. The ITS is used to reconstruct the vertex of the collision in both analyses, and for track reconstruction at mid-rapidity. In addition the Time Projection Chamber (TPC) is used for track reconstruction and particle identification in the dielectron analysis. At forward rapidity, the muon spectrometer is used for triggering and muon track reconstruction. \\}
\indent{At mid-rapidity, events are selected with a minimum bias (MB) trigger, while at forward rapidity the dimuon unlike-sign (MUL) trigger is used. The MB trigger requires a signal in both V0 hodoscopes, located on each side of the interaction point. The MUL trigger requires, in addition to the MB trigger, at least one pair of opposite-sign track segments in the muon spectrometer triggering system, each with a $p_{\rm T}$ above the threshold of the on-line trigger algorithm. The  $p_{\rm T}$ threshold is set to provide 50$\%$ efficiency for muon tracks with $p_{\rm T}$~=~1~GeV/$c$. At mid-rapidity, secondary electrons are rejected by requiring at least one hit in the innermost two layers of the ITS and by using tight selection on the distance of closest approach to the primary vertex. Electrons which form pairs with invariant mass below 50 MeV/$c^{2}$ are removed to further reduce electrons from photon conversion. The J/$\psi$ signal is then extracted in a given $p_{\rm T}$ range, by bin counting in the mass window 2.92 $< m_{e^{+}e^{-}} <$~3.16~GeV/$c^{2}$, from the unlike-sign dielectron (m, $p_{\rm T}$) 2D-distribution after combinatorial background subtraction with the event mixing technique. At forward rapidity, J/$\psi$ candidates are formed by combining pairs of opposite-sign tracks reconstructed in the geometrical acceptance of the muon spectrometer and matching a track segment above the $p_{\rm T}$ threshold in the trigger chambers. In addition, cuts on rapidity  and $p_{\rm T}$ are applied on the J/$\psi$ candidates (2.5 $< y <$ 4, $p_{\rm T} <$ 0.3 GeV/$c$). The J/$\psi$ signal is then extracted from a fit to the dimuon invariant mass spectra. The combinatorial background is either described by an empirical function or subtracted by using an event mixing technique. }

\begin{figure*}[!htpb]
\centering
 \includegraphics[width=0.49\textwidth,clip]{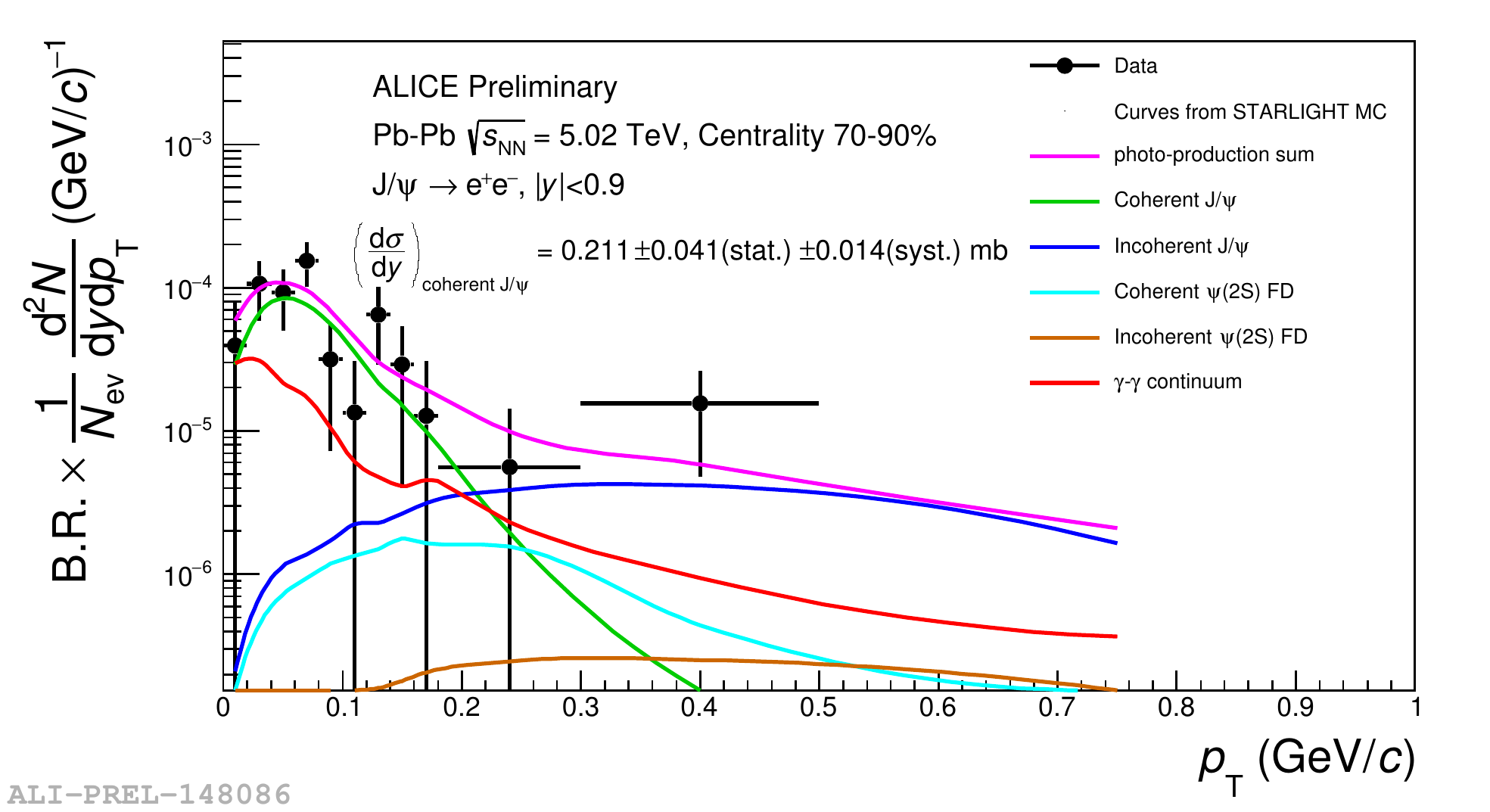}
 \includegraphics[width=0.49\textwidth,clip]{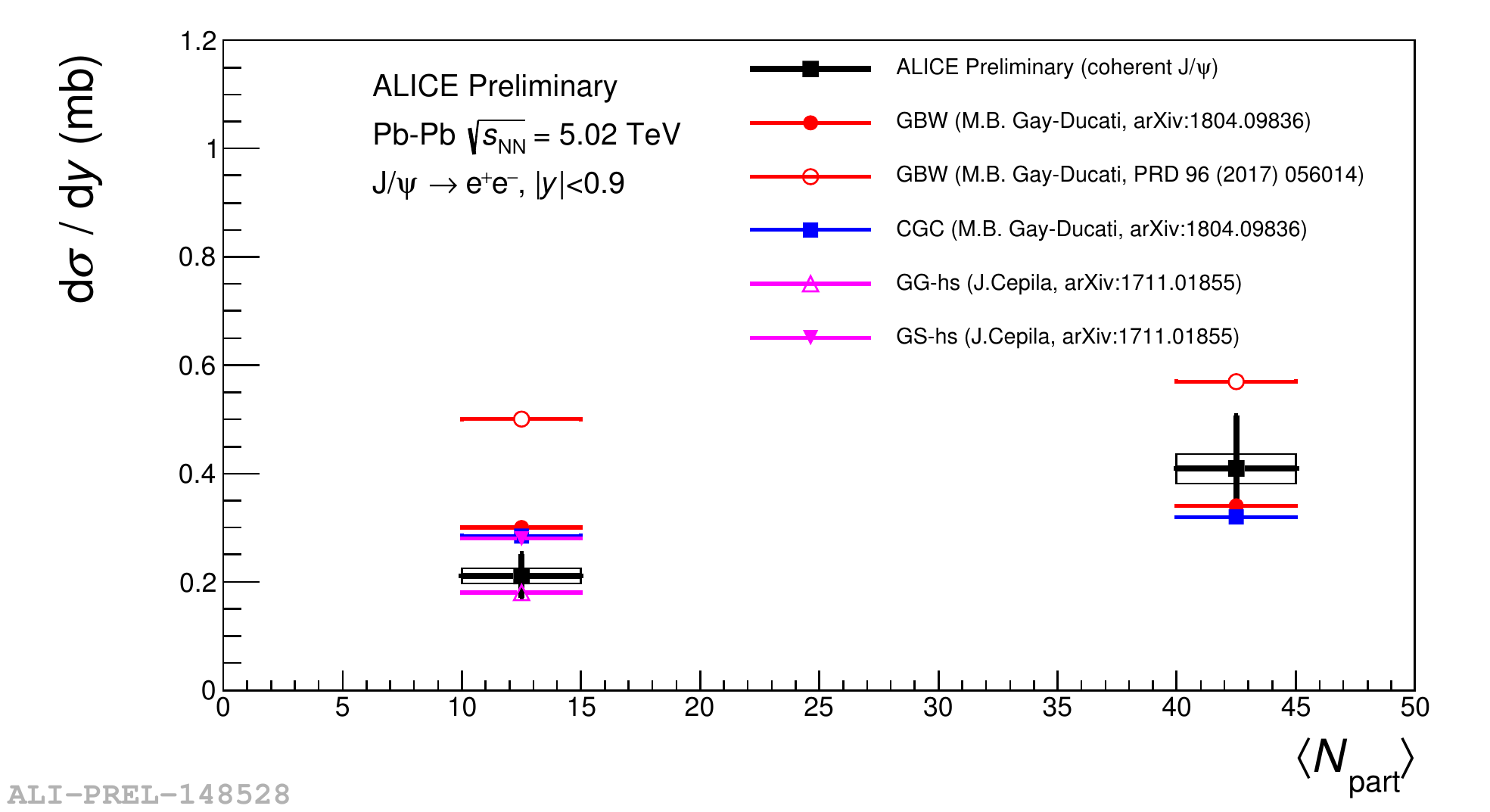}
\caption{\scriptsize{Left: Dielectron yield as a function of $p_{\rm T}$ in the centrality range 70-90$\%$ in Pb-Pb collisions at $\sqrt{s_{\rm NN}}$ = 5.02 TeV, at mid-rapidity. Data are fitted with a cocktail Monte Carlo simulation from the STARLIGHT generator. Right:  Coherent J/$\psi$ photoproduction cross section per unit of rapidity as a function of the number of participating nucleons (corresponding to the centrality ranges 70-90$\%$ and 50-70$\%$), at mid-rapidity. Data are compared to several theoretical models (see text for details).}}
\label{fig-1}       
\end{figure*}

Figure~\ref{fig-1} (left) shows the corrected $p_{\rm T}$-dependent dielectron yield computed with: 
\begin{equation}
B.R. \times \frac{1}{N_{\rm ev}} \frac{d^{2}N(p_{\rm T}) }{dydp_{\rm T}} = \frac{1}{N_{\rm ev}} \frac{N^{\rm raw}(p_{\rm T})}{\Delta y \Delta p_{\rm T} (A \times \epsilon)}
\end{equation}
where $N^{\rm raw}(p_{\rm T})$ is the raw dielectron yield, $ N_{\rm ev}$ is the number of MB events, (A $\times$ $\epsilon$) is the acceptance and efficiency correction assuming coherent J/$\psi$ photoproduction and $y$, $p_{\rm T}$ are the dielectron rapidity and transverse momentum, respectively. Hadronically produced J/$\psi$ are subtracted by fitting the corrected J/$\psi$ $p_{\rm T}$-distribution in the $p_{\rm T}$ range above 0.5 GeV/$c$ and by extrapolating the power-law function down to zero $p_{\rm T}$. The contribution from hadronically produced J/$\psi$ in the range $p_{\rm T} <$~0.2 GeV/$c$ is below 1$\%$.  The corrected $p_{\rm T}$-dependent dielectron yield is fitted with a template including five photoproduction sources obtained from STARLIGHT Monte Carlo (MC) simulations \cite{Klein:2016yzr}: coherently photoproduced J/$\psi$, incoherently photoproduced J/$\psi$, J/$\psi$ from coherently photoproduced $\psi(\rm 2S)$ decays, J/$\psi$ from incoherently photoproduced $\psi(\rm 2S)$ decays and dielectron production from the $\gamma \gamma$ continuum. The relative weights of the sources are fixed to the values obtained in the Pb-Pb UPC measurement which benefits from larger statistics, while the overall scale is left free. The MC cocktail which is able to reproduce the Pb-Pb UPC measurement also provides a good description of the peripheral Pb-Pb data at very-low $p_{\rm T}$, strengthening the hypothesis that coherent photoproduction is at the origin of the observed J/$\psi$ excess. 

\begin{figure*}[tpb]
\vglue -3 true cm
\centering
 \includegraphics[width=0.49\textwidth,clip]{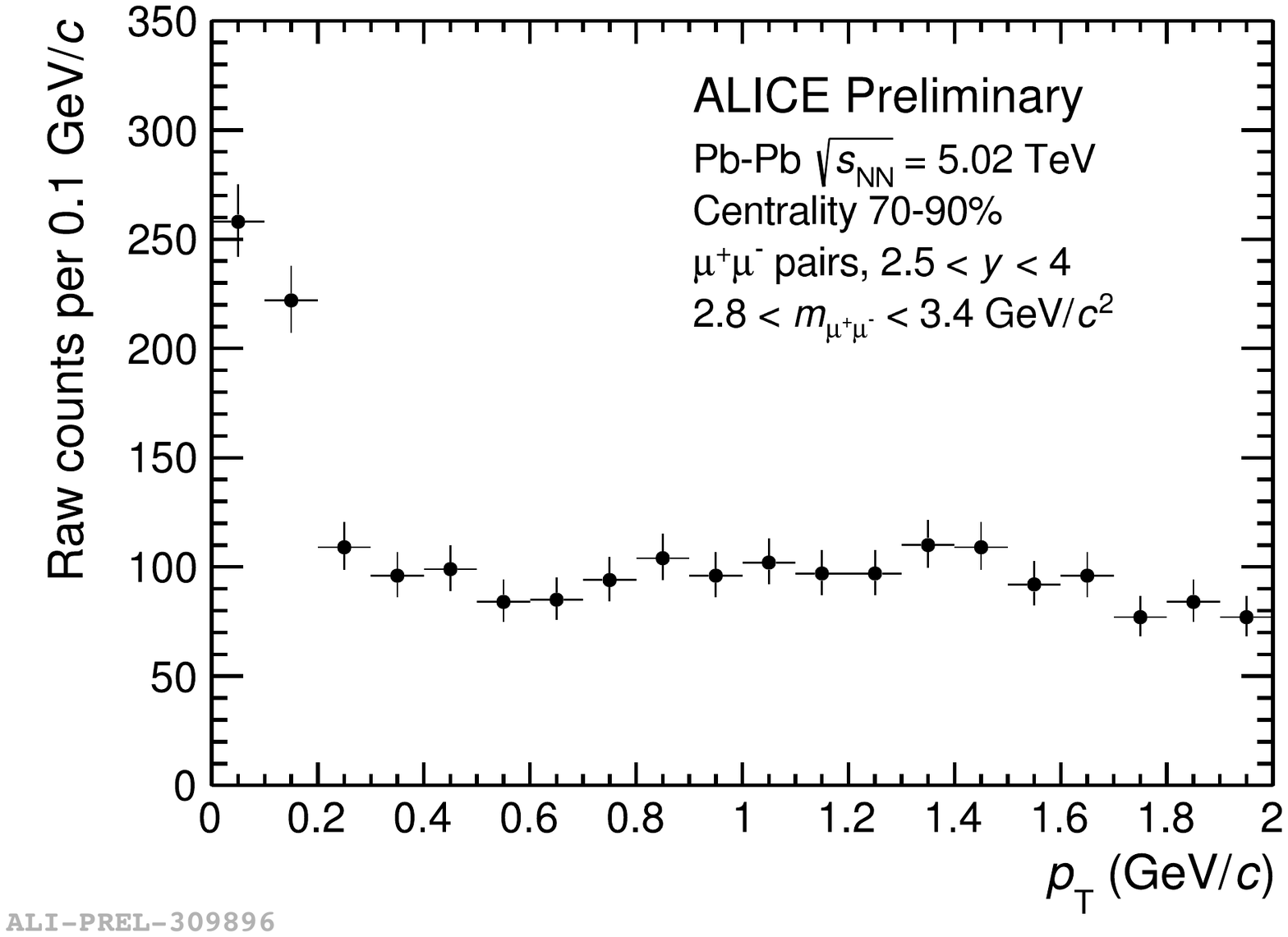}
 \includegraphics[width=0.49\textwidth,clip]{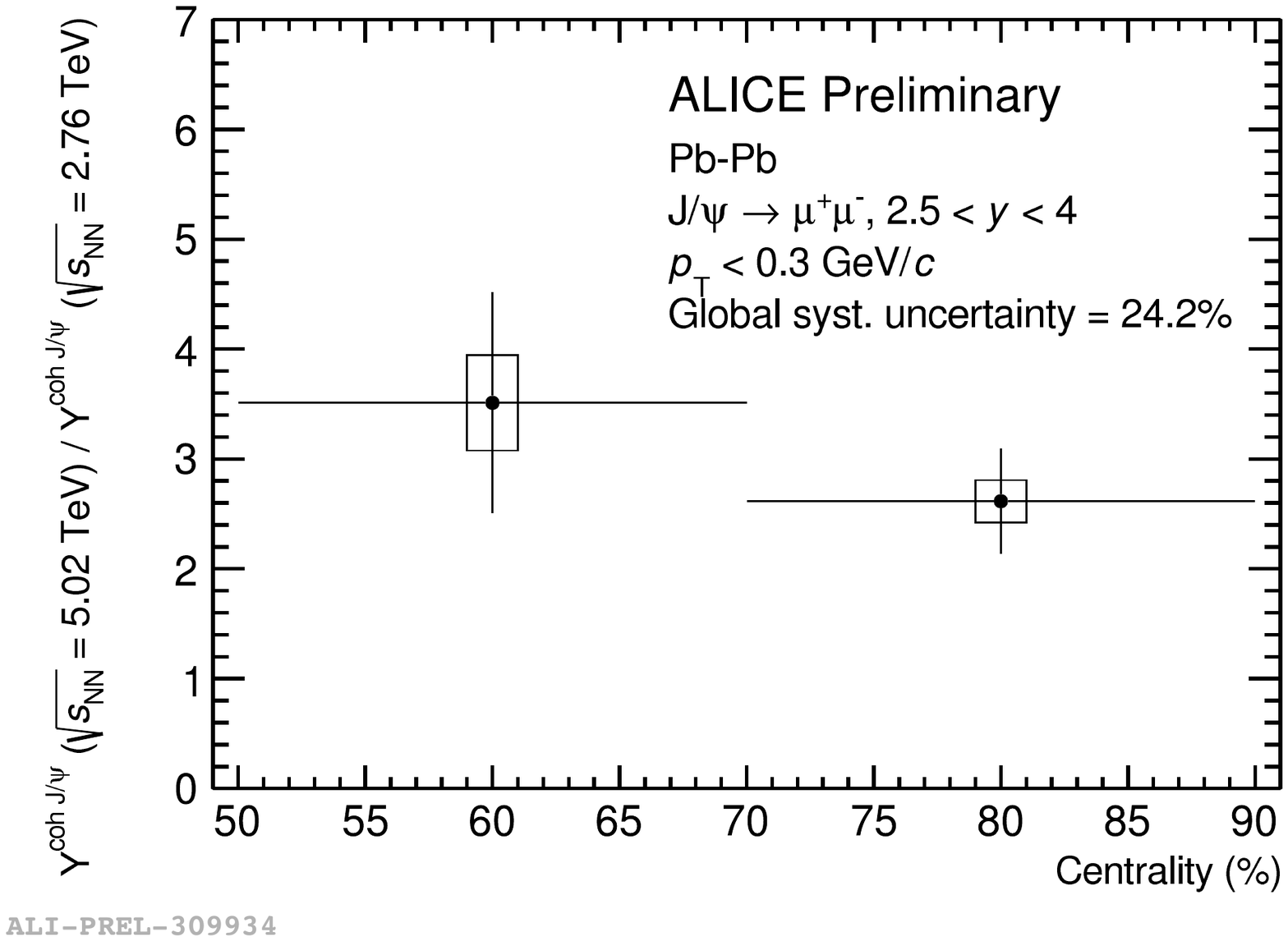}
\vglue -2.5 true cm
\caption{\scriptsize{Left: Uncorrected opposite-sign dimuon $p_{\rm T}$-distribution in the J/$\psi$ mass region (2.8 $< m_{\mu^{+}\mu^{-}} <$ 3.4 GeV/$c^{2}$) at forward rapidity, in the centrality range 70-90$\%$, in Pb-Pb collisions at $\sqrt{s_{\rm NN}}$ = 5.02 TeV. Right: Coherently photoproduced J/$\psi$ yield ratio for two different energies ($\sqrt{s_{\rm NN}}$ = 5.02 TeV over $\sqrt{s_{\rm NN}}$ = 2.76 TeV) as a function of centrality, at forward rapidity in Pb-Pb collisions. J/$\psi$ candidates are selected with a $p_{\rm T}$ below 0.3 GeV/$c$.}}
\label{fig-2}       
\end{figure*}

Figure~\ref{fig-1} (right) compares the coherent J/$\psi$ photoproduction cross section as a function of the number of participating nucleons to several theoretical models. Since the first ALICE measurement of an excess in yield of J/$\psi$ at very-low $p_{\rm T}$ \cite{Adam:2015gba}, several theoretical developments were performed to model the J/$\psi$ photoproduction in peripheral and semi-central Pb-Pb collisions both at $\sqrt{s_{\rm NN}}$~=~2.76 TeV \cite{Zha:2017jch, Klusek-Gawenda:2016oqd} and 5.02 TeV \cite{Ducati:2017sdr, Cepila:2017nef, GayDucati:2018who}. The GBW dipole model incorporates a modification of the photon flux \cite{Ducati:2017sdr} (open circle) and was further refined to also account for a modification of the photonuclear cross section \cite{GayDucati:2018who} (full circle). The CGC model is based on a Color Glass Condensate approach \cite{Ducati:2017sdr}. The GG-hs/GS-hs model is an energy dependent hot spot model, in which the photon-proton interaction is extrapolated to the photon-lead interaction using either a Gribov-Glauber (GG) calculation or a geometric scaling (GS) \cite{Cepila:2017nef}. All models are able to reproduce qualitatively the order of magnitude of the cross section, apart from the GBW model which accounts for the modification of the photon flux only.  Figure \ref{fig-2} (left) shows the raw-$p_{\rm T}$ distribution of opposite-sign dimuons, in the centrality range 70-90$\%$ and invariant-mass range 2.8~$< m_{\mu^{+}\mu^{-}} <$~3.4~GeV/$c^{2}$, at forward rapidity in Pb-Pb collisions at $\sqrt{s_{NN}}$ = 5.02 TeV. It is remarkable that an excess at very-low $p_{\rm T}$ ($p_{\rm T} <$ 0.15 GeV/$c$) is clearly observed in peripheral collisions in the mass range 2.8~$< m_{\mu^{+}\mu^{-}} <$~3.4~GeV/$c^{2}$, where J/$\psi$ contribute to a large fraction of the dimuon yield. This was further confirmed by checking the invariant mass distribution of opposite sign dimuons for $p_{\rm T} <$~0.3~GeV/$c$ which exhibits a clear, almost background free J/$\psi$ signal. The raw J/$\psi$ yield is extracted from a fit to the dimuon invariant mass distribution, in the range $p_{\rm T} < $~0.3 GeV/$c$. The analysis is currently focussing on the two most peripheral centrality intervals (50-70$\%$ and 70-90$\%$). The contribution from hadronically produced J/$\psi$ is subtracted using a data-driven model which uses as inputs the J/$\psi$ $p_{\rm T}$-distribution in pp collisions at $\sqrt{s}$~=~5.02~TeV, the J/$\psi$ nuclear modification factor in Pb-Pb collisions at $\sqrt{s_{\rm NN}}$~=~5.02~TeV and the (A $\times$ $\epsilon$) versus $p_{\rm T}$ of hadronic J/$\psi$ obtained in MC simulations. After the subtraction of the hadronic J/$\psi$ contribution, the significance of the raw J/$\psi$ excess is 14$\sigma$ and 10$\sigma$ in the centrality ranges 70-90$\%$ and 50-70$\%$, respectively. In order to extract the coherent J/$\psi$ yield ($Y^{\rm{coh} \; \rm{J/\psi}}$), the measured excess ($N^{\rm{exc} \; \rm{J/\psi}}$) needs to be corrected by the fraction of incoherently photoproduced J/$\psi$ ($f_{\rm I}$) and the fraction of coherently photoproduced $ \psi(\rm 2S)$ with a J/$\psi$ among the decay products ($f_{\rm D}$), such that:
\begin{equation}
Y^{\rm{coh} \; \rm{J/\psi}} = \frac{\frac{N^{\rm{exc} \; \rm{J/\psi}}}{1+f_{\rm I}+f_{\rm D}}}{B.R. \times (A \times \epsilon) \times N_{\rm ev}}
\end{equation} 
with ($A \times \epsilon$) the acceptance and efficiency for coherently photoproduced J/$\psi$ (assuming transverse polarization), B.R. the branching ratio of the J/$\psi$ in the dimuon decay channel, and N$_{\rm ev}$ the number of minimum bias events. In absence of a measurement of $f_{\rm I}$ and $f_{\rm D}$ in Pb-Pb UPC collisions at $\sqrt{s_{\rm NN}}$~=~5.02~TeV for $p_{\rm T}$~$<$~0.3~GeV/$c$, the ratio of the coherent J/$\psi$ yield between the energies of 5.02 TeV and 2.76 TeV \cite{Adam:2015gba} (Figure \ref{fig-2}, right) has been computed assuming that the fraction $f_{\rm I}$ and $f_{\rm D}$ are energy independent. This assumption is at the origin of the dominant source of systematic uncertainty (20$\%$) on the yield ratio. The coherent J/$\psi$ yield increases by about a factor 2.5 with energy in the centrality range 70-90$\%$ and does not exhibit a strong centrality dependence within uncertainties. The precision on the measurement will improve when the new UPC measurement of $f_{\rm I}$ and $f_{\rm D}$ at $\sqrt{s_{\rm NN}}$ = 5.02 TeV will become available. 

\begin{figure*}[tpb]
\vglue -3 true cm
\centering
 \includegraphics[width=0.49\textwidth,clip]{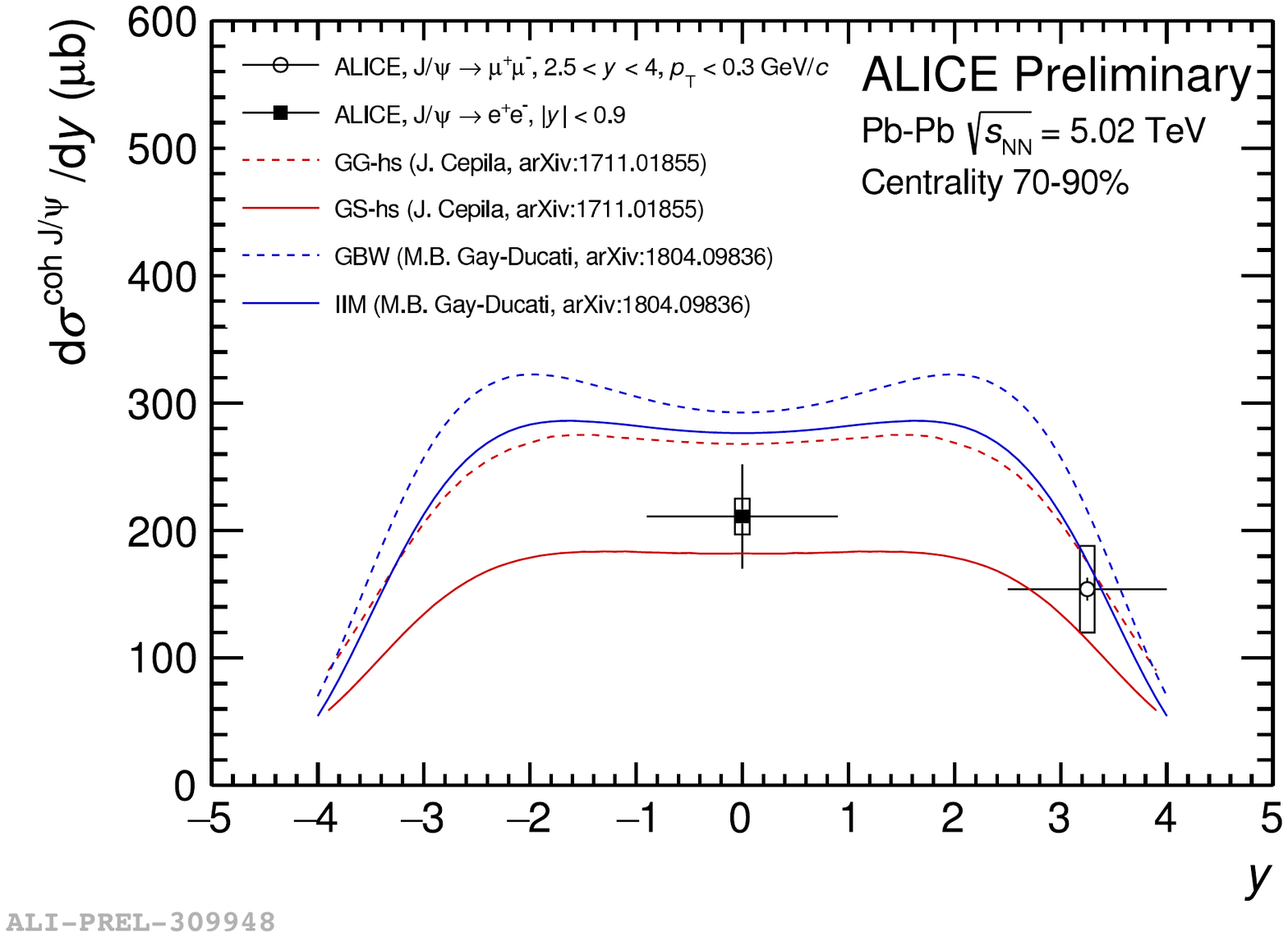}
 \includegraphics[width=0.49\textwidth,clip]{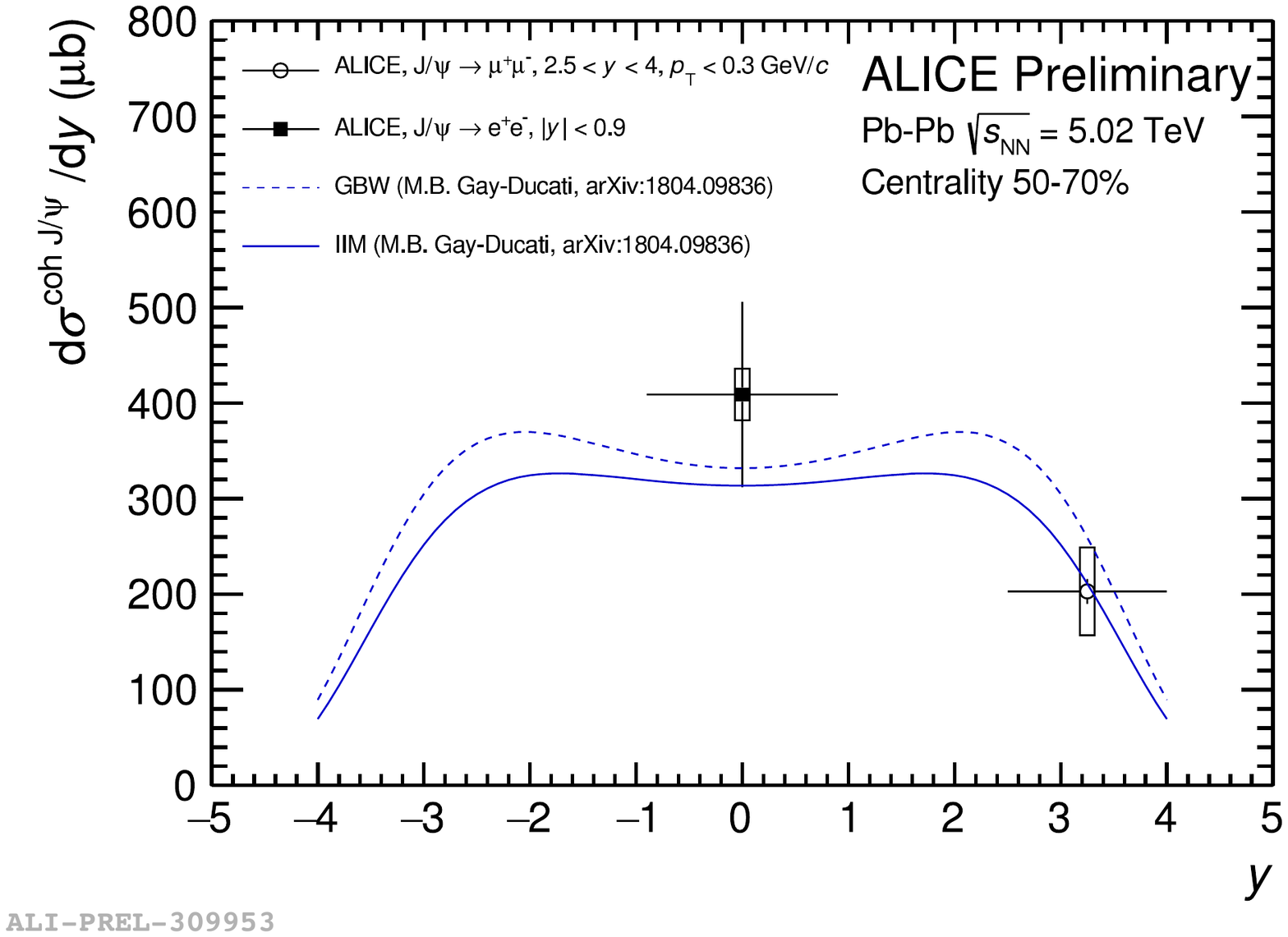}
\vglue -2.5 true cm
\caption{\scriptsize{Coherently photoproduced J/$\psi$ cross section per rapidity unit, at forward and mid-rapidity, in Pb-Pb collisions at $\sqrt{s_{\rm NN}}$~=~5.02~TeV, in the centrality range 70-90$\%$ (left) and 50-70$\%$ (right), compared to model calculations (see text for details).}}
\label{fig-3}       
\end{figure*}

The coherent J/$\psi$ photoproduction cross section at $\sqrt{s_{\rm NN}}$ = 5.02 TeV and forward rapidity is evaluated using the measurement of the J/$\psi$ photoproduction cross section at $\sqrt{s_{\rm NN}}$ = 2.76 TeV in the same centrality range \cite{Adam:2015gba}, the coherent photoproduced J/$\psi$ yield ratio and the hadronic Pb-Pb cross-section ratio between the two energies. The dominant source of systematic uncertainty on the cross-section measurement, like for the yield ratio, comes from the unknown energy dependence of $f_{\rm I}$ and $f_{\rm D}$ (20$\%$). The forward photoproduced J/$\psi$ cross-section measurement at $\sqrt{s_{\rm NN}}$ = 5.02 TeV is shown together with the mid-rapidity measurement in the centrality range 70-90$\%$ (Figure \ref{fig-3}, left), 50-70$\%$ (Figure \ref{fig-3}, right) and compared to models currently available at both rapidities (described in the previous section). The GBW model is the dipole model of \cite{GayDucati:2018who}; the IIM model is the Color Glass Condensate approach of \cite{Ducati:2017sdr}; The GG-hs/GS-hs model is the energy dependent hot spot model of \cite{Cepila:2017nef} with Gribov-Glauber (GG) calculation or geometric scaling (GS). The models that are already able to qualitatively describe the magnitude of the cross section at mid-rapidity are also able to reproduce the magnitude of the forward rapidity measurement in peripheral events.

The first measurement of an excess in the yield of J/$\psi$ at very-low $p_{\rm T}$ in peripheral Pb-Pb collisions at $\sqrt{s_{\rm NN}}$ = 2.76 TeV performed by ALICE has been confirmed with new measurements both at mid and forward rapidity in peripheral Pb-Pb collisions at $\sqrt{s_{\rm NN}}$ = 5.02 TeV. 
Invoking the J/$\psi$ photoproduction mechanism to explain the excess seems further supported by the  $p_{\rm T}$ dependence of the said excess measured at mid-rapidity. Models used to describe Pb-Pb UPC data and modified to account for the nuclear overlap region are able to describe qualitatively the peripheral Pb-Pb data at both rapidities. The new 2018 Pb-Pb data open new experimental perspectives. The evolution with centrality of the J/$\psi$ excess $p_{\rm T}$-distribution at mid-rapidity will allow one to investigate the role of spectator nucleons in the coherence condition. A first J/$\psi$ coherent photoproduction cross-section measurement at forward rapidity in most central events should also be within reach. In addition, a feasibility study of the J/$\psi$ excess polarisation measurement to confirm its photoproduction origin could be conducted. Finally, the first measurement of the photoproduced $\psi(\rm 2S)$ over J/$\psi$ ratio, compared with UPC, will allow one to look for QGP effects on photoproduced charmonia.

\bibliographystyle{utphys_cp} 
\bibliography{skeleton}


\end{document}